\documentclass[showpacs,preprintnumbers,amsmath,amssymb,prl,twocolumn,floatfix,amssymb]{revtex4}
\usepackage[dvips]{graphics,color}
\usepackage{dcolumn}
\usepackage{amsmath,amssymb,bbold,bm}

\begin{document}
\title{Inverse Spin-Galvanic Effect in a Topological-Insulator/Ferromagnet Interface}
\author{Ion Garate$^{1,2}$ and M. Franz$^{1}$}
\affiliation{$^1$Department of Physics and Astronomy, The University of British Columbia, Vancouver, BC V6T 1Z1, Canada}
\affiliation{$^2$Canadian Institute for Advanced Research, Toronto, ON M5G 1Z8, Canada.}
\date{\today}

\begin{abstract}
When a ferromagnet is deposited on the surface of a topological insulator (TI) the topologically protected surface state develops a gap and becomes a 2-dimensional quantum Hall liquid. We demonstrate that the Hall current in such a liquid, induced by an external electric field, can have a large effect on the magnetization dynamics of the ferromagnet by changing the effective anisotropy field.
This change is dissipationless and may be substantial even in weakly spin-orbit coupled ferromagnets.
We study the possibility of dissipationless current-induced magnetization reversal in monolayer-thin, insulating ferromagnets with a soft perpendicular anisotropy and discuss possible applications of this effect.
\end{abstract}
\maketitle

{\em Introduction.---}
Understanding the electric-field control of magnetization and harnessing its technological potential are amongst the most important objectives of spintronics.
Current-induced spin torques can reverse the magnetization of conducting ferromagnets and move magnetic domain walls \cite{spin torque}.
However, the Joule heating generated by transport currents remains a handicap from a practical viewpoint.
An electric field can also reorient the magnetization of insulating compounds with broken inversion symmetry via the magnetoelectric coupling \cite{khomskii2009}.
While they overcome the issue with Joule heating, these multiferroic materials are fewer and more difficult to engineer than common metallic ferromagnets.   
Recently, a novel magnetoelectric effect has been discovered \cite{qi2008a} in topological insulators that are coated with ferromagnetic films.
TIs are bulk insulators with an anomalous band structure that supports topologically robust gapless states at the surfaces \cite{ti}.
These materials are predicted to display a variety of unconventional spintronics effects \cite{ti spintronics}. One unique feature is the  universal quantized topological  magnetoelectric effect \cite{qi2008a}, described by
\begin{equation}
\label{eq:tme}
{\bf M}_{\rm top}=-C_1 \frac{e^2}{2\pi} \bf{E}.
\end{equation}
Here ${\bf M}_{\rm top}$ is the induced magnetization, $C_1$ is a half-integer topological invariant that depends solely on the sign of the time-reversal-symmetry-breaking perturbation, ${\bf E}$ is the applied electric field and $C_1 e^2/2\pi\equiv \sigma_H$ is the  Hall conductance ($\hbar\equiv 1$ throughout).
Unfortunately, the prospects for manipulating the magnetization of real ferromagnets via Eq.~(\ref{eq:tme}) are limited because below the threshold Hall current density ($j_H<1 {\rm A/m}$, see Ref.~[\onlinecite{qhe breakdown}]) the topological magnetic field $B_{\rm top}=\mu_0 M_{\rm top}\lesssim 10^{-6} T$ is very small compared to typical coercive fields ($\gtrsim 0.01 T$) in a ferromagnet.
\begin{figure}
\scalebox{0.33}{\includegraphics{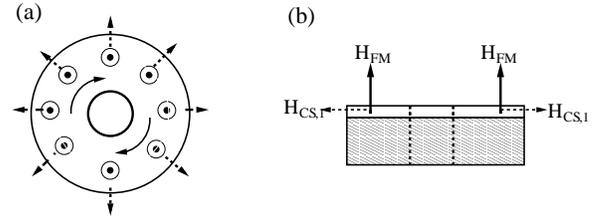}}
\caption{Corbino-disk-shaped TI coated with an ultrathin ferromagnet. (a) Top view: in absence of electric fields, the magnetization of the ferromagnet points outside the page (dotted circles). When a voltage difference is applied between the inner and outer circles, a dissipationless Hall current flows at the interface between the two materials (solid arrows). This current magnetizes the surface states of the TI (inverse spin-galvanic effect) along the radial direction (dashed arrows), thus exerting a spin torque on the magnetization of the ferromagnet.  (b) Cross sectional view: the shaded region is the TI, whereas the unshaded region is the ferromagnetic film. ${\bf H}_{\rm FM}$ is the anisotropy field in electric equilibrium. ${\bf H}_{\rm CS,1}$ is a topological magnetic field proportional to (and parallel to) the applied electric field.
}
 \label{fig:corbino}
\end{figure}

In this Letter we identify a new contribution to the topological magnetoelectric effect, which stems from the current-induced spin polarization of the TI surface states.
Unlike Eq.~(\ref{eq:tme}), this effect depends on material parameters and is not related to Ampere's law;
instead it is the topological counterpart of the inverse spin-galvanic effect found in conducting materials \cite{ivchenko2008}.
In ultrathin (thickness $\lesssim 1 {\rm nm}$) ferromagnetic insulators 
deposited on a surface of TI  (Fig.\ \ref{fig:corbino})
the topological inverse spin-galvanic effect leads to qualitatively stronger spin-torques than Eq.~(\ref{eq:tme}), thus opening the avenue for current-induced control of magnetization without Joule heating.

{\em Functional integral formalism.---}
We begin by reviewing the equation of motion for the magnetization ${\bf M}\equiv M\hat\Omega$ of a classical ferromagnet (in units of 1/volume). 
At low energies the magnitude $M$ is approximately constant and the only dynamical variable is the direction $\hat\Omega=(\Omega_x,\Omega_y,\Omega_z)$.
The time dependence of $\hat\Omega$ may be determined using the functional integral approach \cite{auerbach1994}, which is built on the partition function
\begin{equation}
\label{eq:Z_fm}
Z=Z_0\int D\hat\Omega({\bf x},t)e^{-S_{\rm FM}[\hat\Omega]}.
\end{equation}
$Z_0$ is the partition function corresponding to the equilibrium magnetic configuration $\hat\Omega=\hat\Omega_{\rm eq}$.  
$S_{\rm FM}=S_B-{\cal E}$ is the action for small (quadratic) spin fluctuations, where $S_B=M\int d{\bf x} dt \hat\Omega\cdot(\hat\Omega_{\rm eq}\times\dot{\hat{\Omega}})$ is the Berry phase and ${\cal E}[\hat\Omega]=\int d{\bf x} dt \hat\Omega \chi^{-1}\hat\Omega$ is the micromagnetic energy functional.
$\chi$ is the spin-spin response function.
The semiclassical equation of motion can be derived from $\delta S_{\rm FM}/\delta\hat\Omega=0$,
\begin{equation}
\dot{\hat\Omega}=\hat\Omega_{\rm eq}\times \left(-\frac{1}{M}\frac{\delta{\cal E}}{\delta\hat\Omega}\right).
\end{equation}
A gradient expansion \cite{chi gradient} of $\chi$ yields the venerable Landau-Lifshitz-Gilbert-Slonczewski equation for magnetization dynamics in the presence of damping and transport currents,
\begin{equation}
\label{eq:lls}
\dot{\hat\Omega}=\hat\Omega_{\rm eq}\times{\bf H}-\alpha\hat\Omega_{\rm eq}\times\dot{\hat\Omega}-{\bf v}_{\rm s}\cdot{\nabla}\hat\Omega-\beta\hat\Omega_{\rm eq}\times{\bf v}_{\rm s}\cdot{\nabla}\hat\Omega+...
\end{equation}
${\bf H}$ is an effective magnetic field (in energy units) that includes the anisotropy field, the exchange field as well as external magnetic fields. 
${\bf H}$ determines the easy axis along which the magnetization of a single-domain ferromagnet points in equilibrium.
${\bf v}_{\rm s}$ is the adiabatic spin transfer velocity and is proportional to the transport current.
$\alpha$ and $\beta$ characterize dissipative processes in which energy is transferred from magnetic to non-magnetic (e.g. lattice) degrees of freedom.

{\em Topological effective magnetic field.---}
We now address the magnetization dynamics of an insulating ferromagnet sitting on top of a TI.
The low-energy effective Hamiltonian for the surface states of the TI is \cite{qi2008a,ti}
\begin{equation}
\label{eq:h_ti}
{\cal H}=v_F {\boldsymbol\tau}\cdot({\boldsymbol \pi}\times\hat z)-\Delta{\boldsymbol\tau}\cdot\hat\Omega,
\end{equation}
 where $v_F$ is the Fermi velocity, $\tau^i$ ($i\in\{x,y,z\}$) are Pauli matrices denoting real spin of the surface states, ${\boldsymbol\pi}=-i\nabla-e{\bf A}$, ${\bf A}$ is the electromagnetic vector potential, $\hat z$ is the unit vector normal to the interface between the TI and the ferromagnet and $\Delta$ is the exchange coupling between the surface states and the local moments of the ferromagnet ($\Delta>0$ for ferromagnetic coupling).
We consider a ferromagnet with perpendicular anisotropy ($\Omega_{\rm eq}=\hat z$) so that in equilibrium a gap opens in the energy spectrum of the surface states.
 
The partition function for this composite system is
\begin{equation}
Z=Z_0\int D\hat\Omega({\bf x},t) e^{-S_{\rm FM}[\hat\Omega]}\int D^2\Psi({\bf x},t) e^{-S_{\rm TI}[\bar{\Psi},\Psi,\hat\Omega]},
\end{equation}
where $S_{\rm FM}$ is the ferromagnetic action discussed above and
\begin{equation}\label{hti}
S_{\rm TI}=\int d^2 x dt\bar\Psi\left[\partial_0-\mu -{\cal H}\right]\Psi
\end{equation}
is the action for the surface states. 
$\Psi$ is a fermionic spinor, $\partial_0=\partial_t-eA_0$, $\mu$ is the chemical potential (located in the gap) and $A_0$ is the electrostatic potential.
After rotating the spins by an angle $\pi/2$ around $\hat z$, Eq.\ (\ref{hti})  may be rewritten as $S_{\rm TI}=\int d^2 x dt\bar\psi[\partial_0-\mu-\tilde{\cal H}]\psi$ with
\begin{equation}\label{hd}
 \tilde{\cal H} =v_F \tau^x(\pi_x-e a_x)+v_F\tau^y(\pi_y-e a_y)-\Omega_z\Delta\tau^z,
\end{equation}
where $\psi$ is the rotated fermion field.
In this new basis, ${\bf a}\equiv\Delta/(e v_F)(\hat\Omega\times\hat z)$ appears as an additional contribution to the effective vector potential.
$\Omega_z\Delta$ acts as a mass term.
These massive Dirac fermions may be integrated out in the standard manner \cite{wen2004}, whereby
$Z=\int D\hat\Omega({\bf x},t) e^{-S_{\rm eff}[\hat\Omega]}$.
To second order in $\hat\Omega$ the effective action is
$S_{\rm eff}\simeq S_{\rm FM}+ S_{\rm CS}+S_{\rm EB}$, where
\begin{equation}\label{eq:dS}
S_{\rm CS} = \frac{e^2}{2\pi}C_1\int d^2 x dt \epsilon^{\mu\nu\lambda} {\cal A}_\mu\partial_{\nu} {\cal A}_{\lambda},
\end{equation}
$\vec{\cal A}=(A_0,A_x+a_x, A_y+a_y)$ is the effective vector potential  and $\mu=t,x,y$.
The Chern-Simons action (\ref{eq:dS}) arises in (2+1) dimensional systems with broken time reversal symmetry and nontrivial topology.
The topology of the band structure is encoded in the TKNN \cite{thouless1982} invariant $C_1$. For fermions described by a single Dirac Hamiltonian (\ref{hd}) we have \cite{rosenberg1},
\begin{equation}\label{c1}
C_1=-{1\over 2}{\rm sgn}(\Omega_z\Delta).
\end{equation}

$S_{\rm EB}$  is quadratic in spatial and temporal derivatives of ${\cal A}_\mu$ and encodes the ordinary dielectric/diamagnetic response of the insulator.
Herein we focus on $S_{\rm CS}$, which is first order in the derivatives of ${\cal A}_\mu$ and thus dominates over $S_{\rm EB}$ at long length and time scales. It also produces the effective magnetic field that underlies the inverse spin-galvanic effect which is central to this study.

The semiclassical magnetization dynamics follows from $\delta S_{\rm eff}/\delta\hat\Omega=0$,
\begin{equation}
\dot{\hat\Omega}=\hat\Omega_{\rm eq}\times ({\bf H}_{\rm FM}+{\bf {H}_{\rm CS}})+...
\end{equation} 
where ${\bf H}_{\rm FM}=-\delta S_{\rm FM}/(M\delta\hat\Omega)$ is the effective magnetic field that collects the anisotropy/exchange fields of the isolated ferromagnet and 
\begin{equation}
\label{eq:h_cs}
{\bf H}_{\rm CS}=-\frac{1}{M_{\rm 2D}}\frac{\delta S_{\rm CS}}{\delta\hat\Omega} = -\frac{\sigma_H}{M_{\rm 2D}}\frac{\Delta}{e v_F}\left[{\bf E}+\frac{\Delta}{e v_F}(\hat z\times \dot{\hat\Omega})\right]
\end{equation}
is an additional (topological) contribution to the magnetic field that results from the exchange coupling between the ferromagnet and the TI.
$M_{\rm 2D}$ is the areal magnetization at the interface (in units of 1/area). ${\bf H}_{\rm CS}$ depends on material parameters ($v_F$, $\Delta$, $M_{\rm 2D}$) and is proportional to the Hall conductivity $\sigma_H=C_1 e^2/2\pi$. 
Because the exchange coupling between the surface states and the localized moments of the ferromagnet is local in space, the influence of ${\bf H}_{\rm CS}$ weakens as the thickness of the ferromagnetic film increases.
 
\begin{figure}
\begin{center}
\scalebox{0.33}{\includegraphics{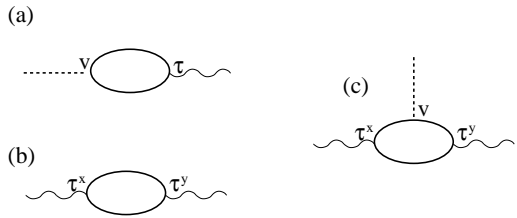}}
\caption{Feynman diagrams for (a) the electric-field-induced magnetization (inverse spin-galvanic effect),
(b) the $xy$ component of the spin-spin response function, (c) the $xy$ component of the spin-spin response function in presence of an electric current (it yields the adiabatic spin transfer torque ${\bf v}_{\rm s}\cdot{\nabla}\hat\Omega$).
The solid straight lines are propagators for massive Dirac quasiparticles (quasiholes).
The solid wavy lines are magnons that couple to the spin operator and the dashed straight lines are photons that couple to the velocity operator.}
\label{fig:bubbles}
\end{center}
\end{figure}

${\bf H}_{\rm CS,1}\equiv -\Delta/(e v_F M_{\rm 2D}) \sigma_H {\bf E}$ can be interpreted as an electric-field induced change of magnetic anisotropy.  
The underlying cause of this effect is that the electric field spin-polarizes the surface states along a direction (${\bf E}/E$) which is misaligned with the equilibrium easy axis ($\hat z$).
We illustrate this point by computing the magnetization induced by a static and uniform electric field:
\begin{equation}
\delta_E M^i_{\rm 2D}\equiv\chi_{M,E}^{ij} E^j,
\end{equation}
where 
\begin{equation}
\chi_{M,E}^{ij} = \lim_{\omega\to 0}\frac{e}{i\omega}\frac{1}{A}\sum_{\bf k}\sum_{n,n'} \tau^i_{n,n'} v^j_{n',n}\frac{f_{{\bf k},n}-f_{{\bf k},n'}}{E_{{\bf k},n'}-E_{{\bf k},n}+\omega}
\end{equation}
is the linear magnetoelectric response function (Fig.~\ref{fig:bubbles}a). 
$n,n'$ are the band indices of the surface states, $E_{{\bf k},n}$ are the band energies, $f_{{\bf k},n}$ are the Fermi distributions, $\tau^i_{n,n'}=\langle n,{\bf k}|\tau^i|n',{\bf k}\rangle$, and $A$ is the area of the interface. 
From Eq.~(\ref{eq:h_ti}), the velocity operator is related to the spin operator via ${\bf v}=\partial{\cal H}/\partial{\bf k}=-v_F {\boldsymbol\tau}\times\hat z$, which
allows us to use the TKNN formula for conductivity \cite{thouless1982} and write
\begin{equation}
\chi_{M,E}^{ij} = -\frac{\sigma_H}{e v_F} \delta^{ij},
\end{equation}
where $\delta^{ij}$ is the Kronecker delta and we have used the fact that the longitudinal conductivity is zero.
Hence $H^i_{\rm CS,1}=(\Delta/M_{\rm 2D}) \delta_E M^i_{\rm 2D}$.
This result is reminiscent of the current-induced effective field in single-domain metallic ferromagnets that belong to the gyrotropic crystal class \cite{manchon_garate}.
Some significant differences between Ref.~[\onlinecite{manchon_garate}] and the present work are that ${\bf H}_{\rm CS,1}$ (i) does not depend on the strength of spin-orbit interactions in the ferromagnet or at the interface (Eq.~(\ref{eq:h_ti}) involves $v_F$ rather than a ``spin-orbit velocity''), (ii) reverses sign when $\Omega_z\to -\Omega_z$ and vanishes when $\Omega_z=0$, (iii) exerts a dissipationless torque provided that the ferromagnet is insulating.

${\bf H}_{\rm CS,2}\equiv-(\sigma_H/M_{\rm 2D})(\Delta/e v_F)^2 \hat z\times\partial_t\hat\Omega$ is associated with the change in the spin response function under a magnetic field (Fig.~\ref{fig:bubbles}b):
\begin{equation}
\label{eq:chi_MB}
\chi_{M,B}^{ij}(q) = \frac{\Delta}{A}\sum_{\bf k}\sum_{n,n'} \tau^i_{n,n'} \tau^j_{n',n}\frac{f_{{\bf k},n}-f_{{\bf k+q},n'}}{E_{{\bf k+q},n'}-E_{{\bf k},n}+\omega},
\end{equation}
where $q=(\omega,{\bf q})$ is the energy/momentum of the magnon and $\tau^i_{n,n'}=\langle n,{\bf k}|\tau^i|n',{\bf k+q}\rangle$. 
At ${\bf q}=0$ we get $\chi_{M,B}^{ij}=(\Delta/e v_F)^2 (-i\omega)\sigma_H \epsilon^{ij}$, where $\epsilon^{xy}=-\epsilon^{yx}=1$ and $\epsilon^{xx}=\epsilon^{yy}=0$.
Thus $H_{\rm CS,2}^i=(\Delta/M_{\rm 2D})\chi_{M,B}^{ij}\Omega_j$ simply increases (if $\Delta>0$) or decreases (if $\Delta<0$) the Berry phase of the isolated ferromagnet\cite{rossier2004}, 
 thereby renormalizing the parameters entering Eq.~(\ref{eq:lls}).


When the magnetization of the ferromagnet is uniform Eq.~(\ref{eq:h_cs}) captures the entire current-induced spin torque for weak electric fields.
In presence of inhomogeneous magnetic textures, one must add the ordinary spin transfer torque. 
The microscopic theory for ${\bf v}_s\cdot{\boldsymbol\nabla}\hat\Omega$ amounts to evaluating the change of the $xy$ spin-spin response function \cite{chi gradient} under an electric field (Fig.~\ref{fig:bubbles}c).
Starting from $\chi^{xy}_{M,B}(q)$, perturbing the matrix elements of the spin operators to first order in {\bf E} 
\cite{fermi factors} and expanding the resulting expression to first order in {\bf q} we find (numerically) that ${\bf v}_s\cdot{\bf q}\propto \Omega_z (E_x q_x-E_y q_y)$.
Furthermore, for realistic parameters the torque exerted by ${\bf H}_{\rm CS}$ is found to dominate over ${\bf v}_s\cdot{\bf q}$ by an ample margin even when $|{\bf q}|\sim{\rm nm}^{-1}$ (note that ${\bf H}_{\rm CS}$  does not vary as $\hat\Omega$ is slightly tilted away from $\hat z$) .

{\em Current-induced magnetization switching.---}
As explained above, ${\bf H}_{\rm CS,1}$ modifies the anisotropy field of the ferromagnet in the presence of a Hall current ${\bf j}_H=\sigma_H \hat z\times {\bf E}$:
\begin{equation}
{\bf H}_{\rm an}=\frac{K}{M_{\rm 2D}}\Omega_z\hat z+\frac{\Delta}{e v_F M_{\rm 2D}}\hat z\times {\bf j}_H,
\end{equation}
where $K$ is the anisotropy energy per unit area for the magnetic ultrathin film in electric equilibrium.
When ${\bf E}=0$ the magnetization of the ferromagnet points along $\hat z$.
After turning on the electric field, the magnetization begins to precess around ${\bf H}_{\rm an}$ and (assisted by the damping) equilibrates along the modified easy axis.
For instance, in a Corbino disk geometry depicted in Fig.~\ref{fig:corbino} the electric field produces a crown shaped magnetization.
Provided that quantum coherence is preserved, this configuration hosts \cite{schutz2003} a circulating spin-current proportional to ${\bf M}(\phi)\times {\bf M}(\phi+\delta\phi)\propto\Omega_z ({\bf j}_H\times \hat z) +O ({\bf E}^2)$, which is radially polarized and persistent (dissipationless). $\phi$ is the azimuthal angle around the disk.

If $j_H\gtrsim e v_F K/\Delta$, $\hat\Omega$ reaches the interface ($\Omega_z=0$) in the course of the precession.  
At that moment, according to Eq.\ (\ref{c1}), $C_1=0$ and hence $\partial_t\hat\Omega=0$; yet this is an unstable fixed point and an infinitesimal in-plane magnetic field suffices to kick the magnetization towards $\Omega_z<0$.
Once this occurs the electric field may be turned off and the magnetization will equilibrate towards $-\hat z$. 
Thus a $180^\circ$ magnetization switching may be completed by combining a dissipationless Hall charge-current with a very small magnetic field. 
Nevertheless, achieving $j_H\gtrsim e v_F K/\Delta$ in real materials presents challenges.
First,  $j_H$ ($E$) cannot be larger than $\sim 1 {\rm A/m}$ ($0.5 {\rm mV/nm}$) because otherwise the dissipationless quantum Hall effect will break down \cite{qhe breakdown}.
Second, we require relatively small coercive fields: $H_{\rm coer}=K/M_{\rm 2D}\lesssim 0.02 T$.  
While such a soft perpendicular anisotropy is inadequate for the magnetic recording industry,
it may find applications in magnetic random access memories and magnetic field sensors \cite{perp anis}.
Third, the thickness of the ferromagnet needs to be comparable to the penetration depth of the Dirac fermions into the ferromagnetic insulator ($\lesssim 1 {\rm nm}$).
While ultrathin films are commonplace in metallic ferromagnets \cite{vaz2008}, insulating ferromagnets such as EuO or EuS present additional experimental difficulties (but see Ref.~\onlinecite{santos2008} for recent progress).
Alternatively, one could electrically manipulate the spin textures caused by magnetic impurities placed on the surface of the TI \cite{liu2009,biswas2009}.  
Using $\Delta= J M_{\rm 2D}$, $j_H=1{\rm A/m}$, $v_F=5\times 10^5 {\rm m/s}$ and $H_{\rm coer}=0.01 T$ we estimate $J\gtrsim 50 {\rm meV nm}^2$ as the condition for magnetization switching. Hence $J/a^2\gtrsim 0.2 {\rm eV}$, where $a\simeq 0.5 {\rm nm}$ is a typical lattice constant for the topological insulator. 
$J/a^2\simeq 0.2 {\rm eV}$ is an {\em a priori} reasonable value \cite{liu2009} for the exchange integral between the localized moments of the ferromagnetic insulator and the surface states of the TI. 
For stronger perpendicular anisotropies (say $H_{\rm coer}\gtrsim 0.05 T$) the exchange integral would need to be of the order of a few eV, and at such strong coupling the surface states of the TI would be altered in a way not captured by Eq.~(\ref{eq:h_ti}).
From the precession frequency $\omega_{\rm prec}\simeq \mu_B H_{\rm an}/\hbar\simeq 1 {\rm GHz}$ we infer switching times of the order of a nanosecond.

There has been some interesting recent work along the lines of the above discussion, albeit in topologically trivial materials \cite{stohr}.
Two salient differences between Ref.~[\onlinecite{stohr}] and the present work are: (i) the microscopic origin of the change in magnetic anisotropy: in our case it is the current-induced spin-polarization of massive Dirac fermions (the topological inverse spin-galvanic effect), whereas Ref.~ \onlinecite{stohr} concentrates on the electrical manipulation of the atomic positions and distortions of the charge distribution; (ii) symmetry of the anisotropy mechanism: in our case it is odd under time reversal (because ${\bf j}_H$ is odd), whereas in Ref.[~\onlinecite{stohr}] it is even under time reversal (because ${\bf E}$ and charge density are even).        

{\em Conclusions.---}
When a ferromagnetic film with perpendicular anisotropy is placed on top of a topological insulator, a Hall-current-induced spin torque arises which modifies the magnetic easy axis.
The origin of this torque can be traced to a topological counterpart of the long-known inverse spin-galvanic effect, and occurs because the electric-field induced magnetization of the surface states in the topological insulator is misaligned with the equilibrium magnetization of the ferromagnet. 
In Corbino disk geometries this effect might be exploited to generate crown-shaped magnetic textures and (for appropriate material parameters) to switch the magnetization of a ferromagnet by $180^\circ$ without Joule heating.

{\em Acknowledgements.---}
We thank I.~Affleck, J.~Folk, A.H.~MacDonald and G.~Sawatzky for helpful comments and questions.
This research has been supported by NSERC  and CIfAR.
I.G. is a CIfAR Junior Fellow.


\begin{thebibliography}{30}
\bibitem{fermi factors} The influence of ${\bf E}$ on the Fermi factors can be neglected because the chemical potential of the surface states lies in the energy gap.
\end{thebibliography}
\end{document}